\input harvmac.tex
\vskip 1.5in
\Title{\vbox{\baselineskip12pt
\hbox to \hsize{\hfill}
\hbox to \hsize{\hfill WITS-CTP-042}}}
{\vbox{
	\centerline{\hbox{New BRST Charges in RNS Superstring Theory
		}}\vskip 5pt
        \centerline{\hbox{and Deformed Pure Spinors
		}} } }
\centerline{Dimitri Polyakov\footnote{$^\dagger$}
{dimitri.polyakov@wits.ac.za}}
\medskip
\centerline{\it National Institute for Theoretical Physics (NITHeP)}
\centerline{\it  and School of Physics}
\centerline{\it University of the Witwatersrand}
\centerline{\it WITS 2050 Johannesburg, South Africa}
\vskip .3in

\centerline {\bf Abstract}
We show that new  BRST charges in RNS superstring theory with 
nonstandard ghost numbers,
constructed in our recent work, can be mapped to deformed pure spinor
(PS) superstring theories, with the nilpotent pure spinor BRST
charge $Q_{PS}=\oint{\lambda^\alpha}d_\alpha$ still retaining its form
but with singular
operator products  between commuting spinor variables $\lambda^\alpha$.
Despite the OPE singularities,
the pure spinor condition $\lambda\gamma^m\lambda=0$ is still
fulfilled in a weak sense, explained in the paper.
The operator product singularities correspond to introducing interactions
between the pure spinors. We conjecture that the leading singularity
orders of the OPE between two interacting pure spinors
is related to the ghost number of the corresponding BRST operator
in RNS formalism. Namely, it is conjectured that
the BRST operators of minimal superconformal
ghost pictures $n>0$ can be mapped to nilpotent
BRST operators in the deformed pure spinor formalism
with the OPE of two
 commuting spinors having a leading singularity order
$\lambda(z)\lambda(w)\sim{O}(z-w)^{-2(n^2+6n+1)}$.
The conjecture is checked explicitly for the first non-trivial
case $n=1$.

\Date{June 2009}
\vfill\eject

\lref\berk{N. Berkovits, JHEP 0801:065(2008)}
\lref\nberk{N. Berkovits, JHEP 0004:018 (2000)}
\lref\nberkk{ N. Berkovits, Phys. Lett. B457 (1999)}
\lref\nberkkk{N.Berkovits, JHEP 0108:026 (2001)}
\lref\green{M.B. Green, J.H. Schwarz, Phys. Lett. B136 (1984) 367}
\lref\fms{D. Friedan, E. Martinec, S. Shenker, Nucl. Phys. B271 (1986) 93}
\lref\sieg{W. Siegel, Nucl. Phys. B263 (1986) 93}
\lref\howe{ P. Howe, Phys. Lett. B258 (1991) 141}
\lref\phowe{ P. Howe, Phys. Lett. B273 (1991) 90}
\lref\verl{H. Verlinde, Phys.Lett. B192:95(1987)}
\lref\bars{I. Bars, Phys. Rev. D59:045019(1999)}
\lref\barss{I. Bars, C. Deliduman, D. Minic, Phys.Rev.D59:125004(1999)}
\lref\barsss{I. Bars, C. Deliduman, D. Minic, Phys.Lett.B457:275-284(1999)}
\lref\lian{B. Lian, G. Zuckerman, Phys.Lett. B254 (1991) 417}
\lref\pol{I. Klebanov, A. M. Polyakov, Mod.Phys.Lett.A6:3273-3281}
\lref\wit{E. Witten, Nucl.Phys.B373:187-213  (1992)}
\lref\grig{M. Grigorescu, math-ph/0007033, Stud. Cercetari Fiz.36:3 (1984)}
\lref\witten{E. Witten,hep-th/0312171, Commun. Math. Phys.252:189  (2004)}
\lref\wb{N. Berkovits, E. Witten, hep-th/0406051, JHEP 0408:009 (2004)}
\lref\zam{A. Zamolodchikov and Al. Zamolodchikov,
Nucl.Phys.B477, 577 (1996)}
\lref\mars{J. Marsden, A. Weinstein, Physica 7D (1983) 305-323}
\lref\arnold{V. I. Arnold,''Geometrie Differentielle des Groupes de Lie'',
Ann. Inst. Fourier, Grenoble 16, 1 (1966),319-361}
\lref\self{D. Polyakov,  Int.J.Mod. Phys A20: 2603-2624 (2005)}
\lref\selff{D. Polyakov, Phys. Rev. D65: 084041 (2002)}
\lref\ampf{S.Gubser,I.Klebanov, A.M.Polyakov,
{ Phys.Lett.B428:105-114}}
\lref\malda{J.Maldacena, Adv.Theor.Math.Phys.2 (1998)
231-252, hep-th/9711200} 
\lref\sellf{D. Polyakov, Int. J. Mod. Phys A20:4001-4020 (2005)}
\lref\selfian{I.I. Kogan, D. Polyakov, Int.J.Mod.PhysA18:1827(2003)}
\lref\doug{M.Douglas et.al. , hep-th/0307195}
\lref\dorn{H. Dorn, H. J. Otto, Nucl. Phys. B429,375 (1994)}
\lref\prakash{J. S. Prakash, 
H. S. Sharatchandra, J.Math.Phys.37:6530-6569 (1996)}
\lref\dress{I. R. Klebanov, I. I. Kogan, A. M.Polyakov,
Phys. Rev. Lett.71:3243-3246 (1993)}
\lref\selfdisc{ D. Polyakov, hep-th/0602209, to appear
in IJMPA}
\lref\wittwist{E. Witten, Comm. Math.Phys.252:189-258 (2004)}
\lref\wittberk{ N. Berkovits, E. Witten, JHEP 0408:009 (2004)}
\lref\cachazo{ F. Cachazo, P. Svrcek, E. Witten, JHEP 0410:074 (2004)}
\lref\barstwist{ I. Bars, M. Picon, Phys.Rev.D73:064033 (2006)}
\lref\barstwistt{ I.Bars, Phys. Rev. D70:104022 (2004)}
\lref\selftwist{ D. Polyakov, Phys.Lett.B611:173 (2005)}
\lref\klebwitt{I. Klebanov, E.Witten, Nucl.Phys.B 556 (1999) 89}
\lref\klebgauge{C. Herzog, I. Klebanov, P. Ouyang, hep-th/0205100}
\lref\ampconf{A. M. Polyakov, Nucl.Phys.B486(1997) 23-33}
\lref\amplib{A. M. Polyakov, hep-th/0407209,in 't Hooft, G. (ed.):
50 years of Yang-Mills theory 311-329}
\lref\wit{E. Witten{ Adv.Theor.Math.Phys.2:253-291,1998}}
\lref\alfa{D. Polyakov, Int.J.Mod.Phys.A22:5301-5323(2007)}
\lref\ghost{D. Polyakov, Int.J.Mod.Phys.A22:2441(2007)}
\lref\sequence{D. Polyakov, arXiv 0905:4858}
\lref\psrns{D. Polyakov, Int.J.Mod.Phys.A24:2677-2687 (2009)}
\lref\progress{D. Polyakov, work in progress}
\lref\brst{{C. Becchi, A. Rouet, R. Stora, Annals Phys.98:287-321 (1976)}}
\lref\alfagauge{D.Polyakov, Int.J.Mod.Phys.A24:113-139 (2009)}

\centerline{\bf Introduction}

Pure spinor formalism
 ~{\berk, \nberk}  is an efficient way of quantizing
Green-Schwarz superstring theory in covariant gauge.
It is also related to RNS description of superstring theory
by mapping the pure spinor variables to those of RNS formalism
(which can be performed with or without introducing non-minimal
fields)
~{\nberkkk, \psrns}.
One advantage of the pure spinor formalism  
is the remarkably simple expression for the 
BRST operator:
\eqn\grav{\eqalign{Q_{PS}=\oint{{dz}\over{2i\pi}}\lambda^\alpha{d_\alpha}
\cr
\alpha=1,...,16
}} 
with the commuting spinor variable $\lambda$ satisfying the 
pure spinor constraint ~{\howe, \phowe, \nberk}:
\eqn\lowen{\lambda\gamma^m\lambda=0}
(where $\gamma^m$ are the $d=10$ gamma-matrices)
and the action given by
\eqn\grav{\eqalign{
S=\int{d^2z}\lbrace{1\over2}\partial{X_m}\bar\partial{X^m}
+p_\alpha\bar\partial\theta^{\alpha}+{\bar{p_\alpha}}\partial\bar\theta^\alpha
+\lambda_\alpha\bar\partial{w^\alpha}+\bar\lambda_\alpha
\partial{\bar{w^\alpha}}\rbrace}}
where $p_\alpha$ is conjugate to $\theta_\alpha$ ~{\sieg}
and $w_\alpha$ is bosonic ghost conjugate to $\lambda^\alpha$.
The action (3) is related to  the standard GS action
by substituting the constraint

\eqn\lowen{d_\alpha=p_\alpha-{1\over2}(\partial{X^m}+
{1\over4}\theta\gamma^m\partial\theta)(\gamma^m\theta)_\alpha=0.}

In our recent  paper ~{\psrns} we observed an isomorphism
 (up to similarity transformation) between  BRST charges in 
pure spinor  and RNS descriptions of superstring theory,
that doesnœôòùt require non-minimal fields ~{\psrns}.
Namely, we observed that if one expresses the commuting spinor variable
$\lambda^\alpha$ 
in terms of RNS fields (up to an overall normalization factor)
as
\eqn\grav{\eqalign{
\lambda_\alpha=be^{{5\over2}\phi-2\chi}\Sigma_\alpha
+2e^{{3\over2}\phi-\chi}\gamma^m_{\alpha\beta}\partial{X_m}
{\tilde\Sigma}^\beta
-2ce^{{1\over2}\phi}\Sigma_\alpha\partial\phi
-4ce^{{1\over2}\phi}\partial\Sigma_\alpha}}
satisfying $\lambda^\alpha=-4\lbrace{Q_{0}},\theta^\alpha\rbrace$
with

\eqn\lowen{
\theta^\alpha=e^{{1\over2}\phi}\Sigma^\alpha}

being the RNS expression for the Green-Schwarz space-time fermionic coordinate 
and 

\eqn\lowen{Q_0=\oint{{dz}\over{2i\pi}}(cT-bc\partial{c}-{1\over2}
\gamma\psi_m\partial{X^m}-{1\over4}b\gamma^2)}

being the BRST charge in RNS theory,
the pure spinor BRST charge (1) is related to RNS BRST charge (7)
by the similarity transformation ~{\psrns}.
Here and elsewhere $\Sigma^\alpha(1,...,16)$ is the 16 component
space-time spinor in RNS formalism, while ${\tilde{\Sigma}}^\alpha$
is used for the 16 component spinor with the opposite GSO parity.

Here $\phi$ and $\chi$ are the bosonized superconformal ghosts
(appearing in the standard bosonization 
relations for the $\beta\gamma$-system ~{\fms}:
$$\gamma=e^{\phi-\chi},\beta=e^{\chi-\phi}\partial\chi=\partial\xi{e^{-\phi}}$$
 while 
the $bc$-system is bosonized in terms of a single free field
$\sigma$: 
$$b=e^{-\sigma},c=e^\sigma.$$

The expression (6) for the $\theta^\alpha$ variable is  canonically conjugate
to the space-time supercurrent 
$j^\alpha=e^{-{1\over2}\phi}{\tilde{\Sigma}}^\alpha$
at picture $-{1\over2}$.
Alternatively, there  also exists a  picture $-{1\over2}$
presentation for $\theta^\alpha$:
\eqn\lowen{\theta^\alpha=ce^{\chi-{3\over2}\phi}\Sigma^\alpha}
which is the canonical conjugate to the space-time supercurrent
at picture ${1\over2}$ (note that, since $\theta^\alpha$ is off-shell
variable, picture-changing operation is only well-defined
for the supercurrent (which worldsheet integral is on-shell) but
not for $\theta^\alpha$ itself; for this reason,
the versions (6), (8) of $\theta^\alpha$ are not directly related
by the picture  changing). We need both the version (6) 
and the version (8) of $\theta^\alpha$ in order to mantain
the picture uniformity in the $d^\alpha$ operator entering
the expression for the pure spinor BRST charge, as
$d_\alpha$ isn't uniform in $\theta$.
The RNS expression for the $d_\alpha$  variable
in the pure spinor BRST charge, obtained from
the expressions (6) and (8) for $\theta_\alpha$,
 is given by ~{\psrns}
\eqn\grav{\eqalign{
d_\alpha=e^{-{1\over2}\phi}{\tilde{\Sigma}}_\alpha
+2c\xi{e^{-{3\over2}\phi}}\gamma^m_{\alpha\beta}\partial{X_m}
\Sigma^\beta-32\partial{c}\partial\xi\xi{e}^{-{5\over2}\phi}
(\partial{\tilde{\Sigma}}_\alpha-{{19}\over6}
{\tilde{\Sigma}}_\alpha\partial\phi)}}
Using the RNS expression (5) for $\lambda^\alpha$ and the expression
(9) for $d_\alpha$, the straightforward computation of
relevant OPEœôòùs was shown to map the pure spinor BRST charge (1)
into the RNS BRST charge (7), up to the similarity transformation:
\eqn\lowen{Q_0^{PS}\rightarrow{e^{-R}}{Q_0^{RNS}}e^R}
where
\eqn\lowen{R=16\oint{{dz}\over{2i\pi}}
\partial{c}c\partial^2\xi\xi{e}^{-2\phi}\partial\chi(z)}
with
\eqn\lowen{\partial^2\xi\xi=2e^{2\chi}\partial\chi}

The pure spinor - RNS correspondence of the BRST charges (1), (7)
contains, however, a subtle point.
While the map (5), (9) formally relates nilpotent charges
(1) and (7), the RNS expression (5) for $\lambda^\alpha$
is not literally a pure spinor variable of the free theory (3).
That is, although the RNS expression (5) reproduces some properties
of $\lambda^\alpha$ ( it is a primary dimension 0 field and a commuting
space-time spinor, given by BRST commutator with  $\theta^\alpha$ (6)),
it only satisfies the pure spinor condition (2) in the weak sense,
described below.
That is, the OPE between  two $\lambda$'s  is actually singular,
with the leading order term being a double pole:
\eqn\grav{\eqalign{\lambda^\alpha(z)\lambda^\beta(w)\sim
{1\over{(z-w)^2}}\partial{b}be^{5\phi-4\chi}\gamma^m_{\alpha\beta}\psi_m
+\partial{b}b{e^{5\phi-4\chi}}\gamma^m_{\alpha\beta}
({1\over4}\partial^2\psi_m+\psi_m{G^{(2)}}(\phi,\chi,\sigma))}}
where $G^{(2)}(\phi,\chi,\sigma)$ is a polynomial in the
bosonized ghost fields of conformal dimension 2.

The appearance of the OPE terms on the right hand side does not, however, 
affect the nilpotence of the BRST charge
(preserving the pure spinor - RNS correspondence (10))
since, as simple analysis shows, the 
 normal ordered term of (13) has a vanishing 
normal product with with $\gamma^m\Pi_m$ 
where $\Pi_m$ is defined by
\eqn\lowen{d_\alpha(z)d_\beta(w)\sim-{{\gamma^m_{\alpha\beta}\Pi_m(w)}
\over{z-w}}+...,}
while the singular term in front of the double pole
has vanishing normal ordering with the $(z-w)$ order term
of the operator product $d_\alpha(z)d_\beta(w)$; for this reason
no simple pole is  produced in the OPE of the $\lambda^\alpha
d_\alpha$ current with itself.
So the weak pure spinor constraint  
\eqn\lowen{\lambda\gamma^m\lambda\approx{0}}
is defined up to the terms with the vanishing normal ordering
with the appropriate OPE terms of $d_\alpha(z)d_\beta(w)$.

Nevertheless, the appearance of the  singular term
in (13)  indicates that although the commuting  spinor variable
(5) still satisfies the  pure spinor condition
(2) in the weak sense, it is not a pure spinor
of a free  theory (3) but of a theory with some 
 interaction introduced. That is, in order to produce
the singularity in the OPE (13), the free action (3) has to be 
deformed with some interaction terms between the $\lambda$ ghosts.
Note that such a deformation would not  generally affect
the Green-Schwarz matter part of the action (3) (
that appears as a result of imposing
 the constraint $d_\alpha=0$) but only the ghost part.

The above observations imply that the pure spinor condition
(2) for $\lambda$ can be relaxed without violating
the nilpotence of the BRST charge (1) : for instance,
the nilpotence still would be preserved in a theory
with interacting pure spinors (with singular operator
products), if a pure spinor condition (2) is satisfied
in the weak  sense so that all the non-vanishing terms
appearing in the OPE 
$lim_{z\rightarrow{w}}\gamma^m_{\alpha\beta}\lambda^\alpha(z)\lambda^\beta
(w)$
 have vanishing normal orderings with 
the appropriate terms of $d_\alpha(z)d_\beta(w)$.
For example,
consider the OPE between $d_\alpha(z)$ 
$d_\beta(w)$ around the midpoint:
\eqn\grav{\eqalign{
d_\alpha(z)d_\beta(w)=-{{{\gamma^m_{\alpha\beta}}\Pi_m^{(1)}
({{z+w}\over2})}\over{z-w}}
+(z-w)^0\gamma^{m_1...m_3}_{\alpha\beta}\Pi_{m_1...m_3}^{(2)}({{z+w}\over2})
\cr
+(z-w){\lbrace}
{\alpha_1}\gamma^m_{\alpha\beta}\Pi^{(3)}_m
+\alpha_2\gamma^{m_1...m_5}
\Pi^{(3)}_{m_1...m_5}\rbrace({{z+w}\over2})}}
where $\alpha_1$ and $\alpha_2$ are some numbers
and suppose that $\lambda$ satisfies the OPE
\eqn\grav{\eqalign{
\lambda_\alpha(z)\lambda_\beta(w)\sim
(z-w)^{-2}\gamma^{m}_{\alpha\beta}A_m({{z+w}\over{2}})
+\gamma^m_{\alpha\beta}B_m({{z+w}\over2})}}

Then the BRST charge is still nilpotent if
$:B^m\Pi_m^{(1)}:=0$ and
 either
$\alpha_1=0$ or
$:A^m\Pi^{(3)}_m:=0$ (other singularities vanish upon
evaluating traces of gamma-matrices).
This precisely is the situation that is realised in
case of the BRST charge of the form (1) in interacting pure spinor theory with
quadratic OPE singularity (corresponding to $Q_0$ in RNS
formalism) and it can be generalised for the interacting
spinors with higher order OPE singularities (see below).

According to the calculation
performed in ~{\psrns}, the theory of interacting pure spinors
with the double pole OPE singularity can be mapped to standard
BRST charge $Q_0$ of RNS theory, up to similarity transformation.
It turns out ~{\sequence} that in RNS theory one can construct,
apart from $Q_0$ (related to superconformal symmetry of the theory),
 a sequence of nilpotent BRST charges $Q_n$,
related to the ground ring of $\alpha$-symmetries, found
in our previous work. The construction is based on 
the local gauge symmetries on the worldsheet in RNS formalism,
associated with the $global$ space-time $\alpha$-transformations,
 generating nonlinear space-time symmetries
in RNS theory, mixing matter and ghost degrees of freedom,
so that the variation of the matter part of the RNS action 
is cancelled by that of the ghost part. 
The generators of these
symmetries can be classified in  terms of the $b-c$ ghost cohomologies
$R_{2n}(n>0)$ ( with $n$ corresponding to the minimal
$superconformal$ ghost number of the ``truncated''
versions of these generators, inducing the s.c. $incomplete$
space-time symmetries ~{\sequence}).
The expressions for the $\alpha$ generators
typically depend on an arbitrary
point $w$  on the worldsheet ; however, this is a weak
dependence since all the $w$ derivatives  of such operators 
are BRST exact. The BRST exact derivative operators,
 generate in turn the $local$ worlsheet gauge symmetries 
 in RNS theory. Identifying the symmetry algebra
of these generators along with the associate generalized
$B$ and $C$-ghosts, it is straightforward to construct
the related nilpotent BRST operators by the standard prescription
~{\brst}.
The obtained BRST charges can be classified in terms
of the ghost cohomologies $H_n$, as was explained in ~{\sequence}.
For example, the BRST charge of $H_1$, constructed in ~{\sequence},
is given by the integral of the current existing at minimal
superconformal picture 1 (that is, it can't be related
to any superconformal ghost pictures below 1) -
unlike the usual BRST charge (7) that
is given
 by the expression at the picture zero and can in principle
be transformed to any other picture since it is an on-shell operator.
Now the natural question to ask is:
just as  the standard BRST operator in RNS theory can be mapped 
to BRST operator in interacting pure spinor theory
with the double pole  (13) in the OPE between $\lambda$'s,
 - are is there any interpretation of new 
 nilpotent BRST operators $Q_n$  in terms of theories with
deformed pure spinors?
In this paper we attempt to show that the answer to this question
is positive - namely, that a sequence of
nilpotent BRST charges found recently in RNS string theory
corresponds to BRST charges
of the form (1) of various deformed pure spinor theories , 
with the deformations
preserving the nilpotence of the charge (1). More precisely,
we shall argue that
 any RNS BRST charge in the cohomology class $H_n$ 
(built on the gauge symmetries associated with the $\alpha$-generators
of $b-c$ cohomology $R_{2n}$)
can be mapped to certain deformed pure spinor theory, so that the
cohomology order $n$ is related to the leading 
 singularity order in the OPE of 2 $\lambda$'s in the interacting
pure spinor theory,
 given by
$2n^2+12n+2$. We shall demonstrate this map precisely for $n=1$
(the first nontrivial case) and conjecture it for higher $n$'s.
The rest of the paper is  organized as follows.
In the section 2 we briefly review the basic properties
of the local worldsheet gauge symmetries, associated with $R_{2n}$
cohomologies and the related BRST charges.
In the section 3 we show that the RNS BRST charge of minimal ghost number 1
(corresponding to the gauge symmetries associated with $R_2$)
can be mapped to deformed pure spinor theory with
the singular OPE of pure spinor variables $\lambda(z_1)$ and $\lambda(z_2)$
with the leading singularity order $(z_1-z_2)^{-16}$. 
In the concluding section we attemt to extrapolate our result
in order to relate the sequence of BRST charges in RNS formalism
to those in deformed pure spinor  formalism with the leading
orders of OPE singularities between pure spinors
corresponding to the cohomology orders of new BRST operators
in RNS formalism.

\centerline{\bf 2. New BRST Operators in  RNS Formalism}

In this section we review the construction of a sequence
of nilpotent BRST charges, found in our recent work ~{\sequence},
also adding some new observations. 
One starts with the ``truncated'' generators of global $\alpha$-symmetries
in space-time ~{\alfa, \ghost, \sequence}.
The ``truncated'' generators  are typically not BRST invariant,
generating ``incomplete'' versions of the space-time symmetries
(incomplete in the sense that they do not involve the $b-c$
sector of the theory).
On the contrary, the ``full version'' generators that are BRST-invariant
and generate the complete space-time symmetry transformations,
including the $b-c$ sector. The truncated generators are, however,
useful as they can be naturally classified in terms of minimal superconformal
ghost numbers they can have. Different ghost number
versions of these generators can be related by  direct or inverse 
picture changing transformations; however for an $\alpha$-symmetry generator
of minimal positive superconformal ghost number $n$ there
exist no ghost picture versions below $n$.
The truncated generators can thus be divided into  classes characterized by
a minimal ghost number $n$.
Structurally, the space-time
symmetry generators characterized by minimal ghost number $n$
 have the form (if taken at minimal positive picture $n$) ~{\sequence,
\alfa,\ghost}:

\eqn\lowen{L^{\alpha_n{I}}=\oint{{dz}\over{2i\pi}}
e^{n\phi}F_{{1\over2}n^2+n+1}^{\alpha_n{I}}(z)}
where  $F_{{1\over2}n^2+n+1}^{\alpha_n{I}}(z)$ are the 
matter primary fields of conformal dimension ${1\over2}n^2+n+1$,
while $\alpha_k(k=1,...n)$ and $I$ are the indices labelling
the generators (see below for explicit explanation for the indices). 
For example, in case of $n=1$ for non-critical 
RNS superstring theory in $d$ dimensions
there are 2 generators $L^{\alpha_1\pm}$ (so $I\equiv(\pm)$)
with 
$$F_{5\over2}^{\alpha_1{+}}\equiv{F(X,\psi)}=
\psi_m\partial^2{X^m}-2\partial\psi_m\partial{X^m}$$
and
$$F_{5\over2}^{\alpha_1{-}}\equiv{F(\varphi,\lambda)}=
\lambda\partial^2{\varphi}-2\partial\lambda\partial{\varphi}$$
where $X^m,\psi^m$ are the space-time cooordinates and their
worldsheet superpartners (RNS fermions) and $\varphi, \lambda$
are  components of the super Liouville field.
Typically, 
in $d$ space-time dimensions the generators of minimal ghost number $n$
include 1 space-time $d$-vector and $n+1$ space-time scalars
(so altogether there are $d+n+1$ space-time symmetry generators
characterized by minimal ghost number $n$).
As it has been shown
that the symmetries induced
by the generators (17) are closely
related to hidden space-time dimensions, it is convenient to organize
the indices $\alpha_k$ and $I$, labelling the generators, as follows.
Namely, it has been shown ~{\alfa, \ghost, \alfagauge, \sequence} 
 that all the symmetry generators 
of the type (17) having minimal ghost numbers from $1$ to $N$
extend the full space-time symmetry group of $d$-dimensional
RNS string theory (including the Liouville direction) from
$SO(2,d)$ to $SO(2,d+N)$, increasing the number of space-time
dimensions by $N$ units.  For each minimal ghost number $n(1\leq{n}\leq{N})$ 
the $d+n+1$ generators, characterized by $n$,
increase the number of space-time dimensions by 1 unit
(extending the symmetry group from $SO(2,d+n-1)$ to $SO(2,d+n)$),
thus each minimal ghost number ``contributes'' a dimension.
Labelling each induced space-time dimension with
$\alpha_n$ (as before, $n$ is a minimal ghost number of the associate
symmetry generators), it is natural to label  the $d+n+1$
generators (17) of minimal ghost number $n$ as $L^{\alpha_n{I}}$
where the index $I=(m,\pm,\alpha_1,...\alpha_{n-1})$
unifies $d$ original space-time dimensions
(labelled by $m=0,...,d-1$) and $n-1$ extra dimensions,
induced by generators with minimal ghost numbers below $n$
(labelled  by $\alpha_1,...,\alpha_{n-1}$). The $\pm$ indices
 are related to the Liouville direction
(e.g. they distinguish between $L^{m+}$ and $L^{m-}$ generators
that induce $d$ translations in $d$-dimensional space time
and $d$ rotations in the Liouville-matter planes respectively).

As has been already pointed out, the space-time generators
 (17) are incomplete: they are not BRST-invariant
(they don't commute with the supercurrent terms of $Q_0$)
and generate truncated (incomplete) version of the $\alpha$-symmetries.
The $BRST$-invariant complete generators of $\alpha$-symmetries
can be obtained from the truncated generators (17)  by using
the $K$-operator procedure, defined as follows ~{\sequence}:
Let $L=\oint{{dz}\over{2i\pi}}V(z)$ be some global symmetry
generator, incomplete (in the sense described above) and not BRST-invariant,
satisfying
\eqn\grav{\eqalign{\lbrack{Q_{brst}},V(z)\rbrack=\partial{U}(z)+W(z)}}
and therefore
\eqn\lowen{\lbrack{Q_{brst}},L{\rbrack}=\oint{{dz}\over{2i\pi}}W(z)}
where $V$ and $W$ are some operators of conformal dimension $1$
and $U$ is some operator of dimension zero.
Introduce the dimension 0 $K$-operator:
\eqn\lowen{K(z)=-4c{e}^{2\chi-2\phi}(z)\equiv{\xi}\Gamma^{-1}(z)}
satisfying
\eqn\lowen{\lbrace{Q_{brst}},K\rbrace=1}
where 
$\xi=e^\chi$ and $\Gamma^{-1}=4c\partial\xi{e^{-2\phi}}$ is the inverse 
picture-changing operator.
Suppose that the $K$-operator (6) has a non-singular OPE with $W(z)$:
\eqn\lowen{K(z_1)W(z_2)\sim{(z_1-z_2)^N}Y(z_2)+O((z_1-z_2)^{N+1})}
where $N\geq{0}$ and $Y$ is some operator of dimension $N+1$.
Then the complete BRST-invariant symmetry generator ${\tilde{L}}$
can be obtained from the incomplete non-invariant symmetry generator
$L$ by the following transformation:
\eqn\grav{\eqalign{
L\rightarrow{\tilde{L}}(w)=L+{{(-1)^N}\over{N!}}
\oint{{dz}\over{2i\pi}}(z-w)^N:K\partial^N{W}:(z)
\cr
+{1\over{{N!}}}\oint{{dz}\over{2i\pi}}\partial_z^{N+1}{\lbrack}
(z-w)^N{K}(z)\rbrack{K}\lbrace{Q_{brst}},U\rbrace}}
where $w$ is some arbitrary point on the worldsheet.
It is straightforward to check the invariance
of ${\tilde{L}}$ by using some partial integration along with
the relation (7) as well as the obvious identity
\eqn\lowen{\lbrace{Q_{brst}},W(z)\rbrace=
-\partial(\lbrace{Q_{brst}},U(z)\rbrace)}
that follows directly from (4).
The corrected invariant ${\tilde{L}}$-generators
are then typically of the form 
\eqn\lowen{{\tilde{L}}(w)=\oint{{dz}\over{2i\pi}}(z-w)^N{\tilde{V}}_{N+1}(z)}
(see the rest of the paper for the concrete examples)
with the conformal dimension $N+1$ operator ${\tilde{V}}_{N+1}(z)$
in the integrand satisfying
\eqn\lowen{{\lbrack}Q_{brst},{\tilde{V}}_{N+1}(z)\rbrack
=\partial^{N+1}{\tilde{U}}_0(z)}
where ${\tilde{U}}_0$ is some operator of conformal dimension zero.
Applying the $K$-transformation (23) to the symmetry generators (17)
one finds that the BRST-invariant expressions  for the
$\alpha$-symmetry generators
(inducing the full version of the space-time symmetries) are given by:
\eqn\grav{\eqalign{L^{\alpha_n{I}}\rightarrow
{\tilde{L}}^{\alpha_n{I}}(w)={1\over{(2n)!}}\oint{{dz}\over{2i\pi}}(z-w)^{2n}
{\lbrace}e^{n\phi}P^{(2n)}_{2\phi-2\chi-\sigma}
F_{{1\over2}n^2+n+1}^{\alpha_n{I}}\cr
-4c\xi{e}^{(n-1)\phi}({1\over{(n+1)!}}P^{(n+1)}_{\phi-\chi}
L_{{1\over2}n^2+n+{1\over2}}^{\alpha_n{I}}
+{{{f}(n)}\over{n!}}P^{(n+1)}_{\phi-\chi}{\partial}
L_{{1\over2}n^2+n+{1\over2}}^{\alpha_n{I}}
\cr
+\sum_{m=0}^{n-1}{1\over{m!(n-m-1)!}}
P^{(n-1-m)}_{\phi-\chi}\partial^m{G}
F_{{1\over2}n^2+n+1}^{\alpha_n{I}})
+g(n)\partial{c}c\partial\xi\xi{e^{(n-2)\phi}}
F_{{1\over2}n^2+n+1}^{\alpha_n{I}}\rbrace}}
Here $G=-{1\over2}\psi_m\partial{X^m}$ is the matter part of the worldsheet
supercurrent,$L_{{1\over2}n^2+n+{1\over2}}^{\alpha_n{I}}$
is the worldsheet superpartner of 
$F_{{1\over2}n^2+n+{1}}^{\alpha_n{I}}$
satisfying the OPE
\eqn\lowen{G(z)F_{{1\over2}n^2+n+{1}}^{\alpha_n{I}}(w)
={{L_{{1\over2}n^2+n+{1\over2}}^{\alpha_n{I}}}\over{(z-w)^2}}
+{{f(n)\partial{L_{{1\over2}n^2+n+{1\over2}}^{\alpha_n{I}}}}\over{z-w}}
+...}
and $f(n),g(n)$ are some numbers;
for the $n=1,2,3$ cases the value of $f(n)$ was computed 
to be equal to ${1\over4}$, while $g(1)=24,g(2)=20,g(3)=7$.
For $n>3$ cases the computation of the values of
$f(n)$ and $g(n)$ is more complicated, requiring lengthy evaluations
of cumbersome OPEs, although in principle it can be done explicitly.
The complete invariant $\alpha$-symmetry generators can be classified
in terms of $b-c$ ghost cohomologies $R_{2n}$ (where $n$
refer to the minimal $superconformal$ ghost numbers of the
truncated symmetry generators prior to the $K$-transformation (23))
which are defined as follows (see also ~{\sequence}).
One starts with defining the notion of a $b-c$ picture
for physical vertex operators, which is the generalization
of a usual superconformal $\beta-\gamma$ picture. 
We define a physical operator ${\tilde{L}}(w)$ to have a $b-c$ picture $n$
if it is represented in the form
$$L(w)=\oint{{dz}\over{2i\pi}}(z-w)^n{V_{n+1}}(z)$$
where $V_{n+1}$ is some operator of conformal dimension $n+1$
satisfying 
\eqn\lowen{
{\lbrack}Q_0,V_{n+1}\rbrack=\partial^{n+1}U_0} 
and $U_0$ is some operator of dimension $0$.
For example, the pictures $-1$ and $0$ reproduce the familiar
versions of the ``unintegrated''  and ``integrated'' vertex operators
(e.g. a momentum zero photon operator is given by
$c\partial{X^m}+{1\over2}\gamma\psi^m$ at picture $-1$
and $\oint\partial{X^m}$ at picture zero).
In addition, there are no nontrivial $b-c$ ghost pictures below $-1$.
Just like the usual superconformal pictures
can be raised by the transformation with
the dimension zero
picture-changing operator $\Gamma=\delta(\beta)\delta(S)
\equiv:{e^\phi}S:$ (obtained by the integration
over the fermionic supermoduli of gravitini in 
functional integrals for scattering amplitudes with
S being the full matter$+$ ghost supercurrent),
the $b-c$-pictures are raised by the 
dimension zero  invariant $Z$-operator,
that also can be obtained from  functional
integrals for RNS scattering amplitudes by integration
over the $bosonic$ moduli of the worldsheet metric ~{\sellf}:
\eqn\lowen{
Z(w)=\delta(b)\delta({T})\equiv{b}\delta(T)(w)
=\oint{{dz}\over{2i\pi}}(z-w)^3(bT+4c\partial\xi\xi{e^{-2\phi}}T^2)(z)}
where $T$ is the full matter$+$ghost stress-energy tensor.
Just as the standard picture changing operator can be written as the 
BRST commutator outside the small Hilbert space:
$\Gamma=\lbrace{Q_0},\xi\rbrace$, the $Z$-operator (30) is also given by the
BRST commutator outside the small Hilbert space:
\eqn\lowen{
Z(w)=-{\lbrack}Q_0,\oint{{dz}\over{2i\pi}}
(z-w)^4\partial\xi\xi{e^{-2\phi}}{T^2}(z)\rbrack}
Typically, the action of $Z$ on a physical vertex operators
at $b-c$ ghost picture $n$
is given by (after integrating out total derivatives):
\eqn\lowen{Z(w){\tilde{L}}_{\lbrack{n}\rbrack}(w)
\equiv{Z}(w)\oint{{dz}\over{2i\pi}}(z-w)^n{V_{n+1}}(z)
=
{\tilde{L}}_{\lbrack{n+1}\rbrack}(w)\equiv
\oint{{dz}\over{2i\pi}}(z-w)^{n+1}{V_{n+2}}(z)}
where $V_{n+1}$ and $V_{n+2}$ both satisfy (29)
(generally, with the different $U_0$'s).
For example, acting with $Z$ on elementary 
vertex operators (such as photon) at $b-c$ picture $-1$
(known as ``unintegrated'' vertices)  one obtains
vertex operators at $b-c$ picture zero (known as ``integrated''
vertices)
One can also define the $Z^{-1}$ operator 
\eqn\lowen{Z^{-1}=\lbrack{Q_0},\xi(\partial{c}-c\partial\phi)\rbrack}
formally satisfying
\eqn\lowen{\lbrace{Z^{-1}},Z\rbrace=:\Gamma:+\lbrack{Q_0},...\rbrack}

Having made all these definitions, we are now prepared to define
the $b-c$ ghost cohomologies $R_N$. The definition is quite similar
to that of the superconformal ($\beta-\gamma$) ghost cohomologies
described in ~{\sequence} and other works. 
The $b-c$ ghost cohomology $R_N$ consists
of physical (BRST-invariant and nontrivial) vertex operators
, $violating$ the equivalence of the $b-c$ ghost pictures
(defined above), that exist at $minimal$  $b-c$ ghost picture
$N>{-1}$ and $cannot$ be related  to 
$b-c$ pictures less than $N$ by any $Z$-transformation;
the $Z$-transformations can, however, relate it 
to $b-c$ pictures higher than $N$ ($N+1,N+2,...$)
so the elements of $R_{2N}$  exist at pictures $N$ and above,
but not below $N$.
As it is clear from the above definitions,
the complete BRST-invariant $\alpha$-symmetry generators
${\tilde{L}}^{\alpha_n{I}}$ obtained in (27) are the elements
of $R_{2n}$. That is, the truncated non-invariant 
global symmetry generators
$L^{\alpha_n{I}}$ (17) inducing incomplete symmetry transformations
classified by the minimal $superconformal$ ghost number $n$,
become the elements of the $b-c$ ghost cohomology $R_{2n}$
as a result of the $K$-transformation, defined in (23).

The peculiar property of the full symmetry generators (27)
is that, while they induce global nonlinear symmetries 
in space-time, they also depend on an arbitrary point $w$ on the worldsheet,
except for the trivial case $N=0$. The $N=0$ case is realised, for example,
when the $K$-transformation (23) is applied to the incomplete
(and BRST non-invariant) to the space-time rotation generator
$L^{mn}=\oint{{dz}\over{2i\pi}}\psi^m\psi^n$
which induces space-time rotational symmetries for RNS fermions, but not
for RNS bosons; when the $K$-transformation is applied to $L^{mn}$,
the obtained operator ${\tilde{L}}^{mn}$ - BRST-invariant and complete
- generates the space-time rotation for the full set of the matter fields
(up to picture-changing for $X$'s)). The $K$-transformation
applied to the truncated $\alpha$-generators requires, however,
$N=2n$, hence the full invariant generators (27) appear to 
manifestly depend on $w$.
This ambiguity, however, is resolved if we note that
all the  $2n$ non-vanishing derivatives in $w$
 of the complete $\alpha$-generators
$\partial^{k}{\tilde{L}}^{\alpha_n{I}}(w)(k=1,...,2n)$
are BRST-exact:
\eqn\grav{\eqalign{\partial^{k}{\tilde{L}}^{\alpha_n{I}}(w)
=\lbrace{Q_0},\partial^{k-1}(b_{-1}\partial^{k}{\tilde{L}}^{\alpha_n{I}}(w))
\rbrace}} 
In particular, in the $k=1$ case we have
\eqn\lowen{\partial{\tilde{L}}^{\alpha_n{I}}(w)
=\lbrace{Q_0},b_{-1}{\tilde{L}}^{\alpha_n{I}}(w)
\rbrace.}
implying that the left hand side of (36)
(expression of conformal dimension $1$ 
with  the integrand of conformal dimension 2)
is the analogue of the worldsheet integral of the stress-energy tensor
$\oint{T}$, given by the BRST commutator with the worldsheet
integral of generalized
$b$-ghost
\eqn\lowen{\oint{B^{\alpha_n{I}}}(w)=
b_{-1}{\tilde{L}}^{\alpha_n{I}}(w)}
where $b_{-1}$ is the worldsheet integral
of the standard $b$-ghost field:
$b_{-1}=\oint{{dz}\over{2i\pi}}b(z)$.
Next, remarkably, it can be shown ~{\sequence} that
the derivatives of the full $global$ space-time $\alpha$-symmetry generators
${\tilde{L}}^{\alpha_n{I}}(w)$ generate $local$ 
gauge  symmetries on the worldsheet - just like the stress-energy tensor
and its derivatives, given by the anticommutator of the BRST charge
with the $b$-ghost and its derivatives, generate the superconformal
symmetries on the worldsheet. The difference, however, is that
while the superconformal symmetry is infinite-dimensional
(particularly generated by the stress tensor $T$,
the worldsheet supercurrent  and  the infinite number of their derivatives),
the gauge symmetries induced by the derivatives of the $\alpha$-generators
(27) are finite-dimensional, since the number of the non-vanishing
derivatives is finite
(equal to $2n$ for each ${\tilde{L}}^{\alpha_n{I}}$). 
Nevertherless, it turns out that the 
algebra of the gauge symmetries generated by the derivatives of
${\tilde{L}}^{\alpha_n{I}}(w)$ does have a conformal-like structure,
reminiscent of the ``truncated'' finite-dimensional Virasoro algebra
~{\sequence}. Namely, defining
\eqn\lowen{L_{k}^{\alpha_n{I}}=\partial^k{\tilde{L}}^{\alpha_n{I}}}
and rescaling
\eqn\lowen{T_{k}^{\alpha_n{I}}={{L_{k}^{\alpha_n{I}}}\over{(n-k)!}}}
it can be checked explicitly (for $n=1,2,3$ and extrapolated
for higher values of $n$)
that the properly normalized local gauge symmetry generators
$T_{k}^{\alpha_n{I}} (k=1,...,2n)$ satisfy the following commutation relations:
\eqn\grav{\eqalign{{\lbrack}T_{k_1}^{\alpha_{n_1}{I}},
T_{k_2}^{\alpha_{n_2}{J}}\rbrack=
(k_1-k_2){\lbrace}
\theta(max(2n_1,2n_2)-k_1-k_2)\eta^{IJ}T_{k_1+k_2}^{\alpha_{n_1}\alpha_{n_2}}
\cr
+\theta(max(2N(I),2N(J))-k_1-k_2)
\delta^{\alpha_{n_1}\alpha_{n_2}}T_{k_1+k_2}^{IJ}
\cr
+\theta(max(2n_2,2N(I))-k_1-k_2)
\delta^{\alpha_{n_1}{J}}T_{k_1+k_2}^{\alpha_{n_2}{I}}
\cr
-\theta(max(2n_1,2N(J))-k_1-k_2)
\delta^{\alpha_{n_2}{I}}T_{k_1+k_2}^{\alpha_{n_1}J}{\rbrace}}}
where $\theta(n)$ is a usual step function (taken at discrete
values of the argument), equal to 1 for $n\geq{0}$and 0 for $n<0$;
the function $max(m,n)=m$ if $m>n$ and $max(m,n)=n$ otherwise;
finally, $2N(I)$ is the value of the cohomology order associated with
the index $I$ (i.e. $N(I)=k$ if $I=\alpha_k$ and $N(I)=0$ if
$I$ stands for $+,-$ or a $d$-dimensional space-time index $m$).
The step function factors in the commutators (40) 
ensure that for each of the terms 
on the right hand side of (40) the derivative order $k_1+k_2$
of the underlying global $\alpha$-symmetry generator $L^{\alpha_N{I}}$
is less or equal to  its $b-c$ cohomology order $2N$;
since all the higher order derivatives
$L_{k}^{\alpha_{N}I}\equiv\partial^{k}{\tilde{L}}^{\alpha_{N}I}$
vanish identically if $k>2N$ (so that each global $\alpha$-
generator of $R_{2N}$ gives rise to $2N$ local gauge symmetry generators).
So structurally the commutation relations (40) take the form
\eqn\lowen{{\lbrack}
L^{{\underline{M}}}_m,L^{{\underline{N}}}_n{\rbrack}
=(m-n)f^{{\underline{M}}{\underline{N}}}_{{\underline{P}}}
L^{{\underline{P}}}_{m+n}}
(where for convenience
 the underlined indices unify $\alpha_i$ and $I$
in the gauge symmetry generators)
provided that
$m+n$ is less or equal to the R-cohomology order 
of ${\tilde{L}}^{{\underline{P}}}$.
Here $f^{{\underline{M}}{\underline{N}}}_{{\underline{P}}}$
are the structure constants inherited from the global symmetry
algebra of the $\alpha$-generators while the origin
of the $(m-n)$ factor
is Virasoro-type, related to the local worldsheet properties
of the generators (38), (39).

Given the gauge symmetry
generators (35), it is straightforward to check that
the generalized $B$-ghost fields (37)
$\oint{B}^{\alpha_n{I}}=b_{-1}{\tilde{L}}^{\alpha_n{I}}(w)$
are in the adjoint of the gauge symmetries (40).
Thus the only components missing for the construction
of nilpotent BRST charge are the generalized $C$-ghosts
that must be canonical congugates of the $B$-ghosts.
Thus in order to construct the $C$-ghosts one has to identify
local primary fields of dimension $-1$ satisfying
\eqn\lowen{\lbrace\oint{B}^{\alpha_n{I}},C^{\alpha_n{I}}\rbrace=1}
(in full analogy with the usual $b-c$ ghosts).
Such objects have been constructed explicitly for the $n=1,2,3$ cases.
The explicit expressions are given by ~{\sequence}:
\eqn\grav{\eqalign{C^{{\alpha_1}I}
=:e^\phi{G}{e}^{2\phi-\chi}L_{2}^{\alpha_1{I}}:+\lbrack{{\tilde{Q}}_0,
be^{3\phi-\chi}F_{5\over2}^{\alpha_1{I}}P^{(1)}_{\phi-\chi-{3\over4}\sigma}
\rbrack}}}
for the gauge symmetries  derived from
the $\alpha$-generators of $R_2$,
\eqn\grav{\eqalign{C^{\alpha_2{I}}
=e^\phi{G}e^{3\phi-\chi}L^{\alpha_2{I}}_{{9\over2}}
P^{(1)}_{\phi-{{10}\over3}\chi}
+\lbrace{{\tilde{Q}}_0},\partial^2{b}\partial{b}b{e^{5\phi-2\chi}}
{L^{\alpha_2{I}}_{{9\over2}}}
\rbrace}}
for the gauge symmetries derived from
the $\alpha$-generators of $R_4$ and
\eqn\grav{\eqalign{C^{\alpha_3{I}}=
:e^\phi{G}e^{4\phi-\chi}
(P^{(1)}_{\phi-3\chi}\partial{L_8^{\alpha_3{I}}}^{\alpha_3{I}}+{4\over5}
L_8^{\alpha_3{I}}
(\partial{P^{(1)}_{\phi-3\chi}}+P^{(1)}_{\phi-3\chi}P^{(1)}_{4\phi-\chi})):
\cr
+\lbrace{\tilde{Q}}_0,\partial^4{b}\partial^3{b}\partial^2{b}\partial{b}{b}
e^{7\phi-3\chi}(\partial{P^{(1)}_{\phi-{9\over4}\chi}}
+
{P^{(1)}_{\phi-{9\over4}\chi}}P^{(1)}_{7\phi-3\chi-7\sigma})
F_{ij({{17}\over2})}^{matter}\rbrace}}
for the gauge symmetries derived from
the $\alpha$-generators of $R_6$.
Here ${\tilde{Q}}_0=Q_0-\oint{{dz}\over{2i\pi}}(cT-bc\partial{c})$
stands for the supercurrent part of the standard BRST charge
and $G=-{1\over2}\psi_m\partial{X^m}$ is the matter part of the worldsheet
supercurrent.
It should be noted that $C^{\alpha_1{I}},C^{\alpha_2{I}},C^{\alpha_3{I}}$
ghost fields exist at the minimal superconformal
picture 2,3 and 4 respectively; therefore in order to satisfy 
the canonical relations (42) the
$\oint{B^{\alpha_i{I}}}$ ghost fields ($i=1,2,3$) have to be 
picture transformed to superconformal pictures $-2,-3$ and $-4$
respectively by using the inverse picture changing.
Note that, even though $\oint{B^{\alpha_i{I}}}$ are not on-shell
operators, picture-changing transformations (both direct and inverse)
are still well-defined for them
since $L_1^{\alpha_i{I}}=\lbrace{Q_0},\oint{B^{\alpha_i{I}}}\rbrace
\equiv\lbrace{Q_0},b_{-1}{\tilde{L}}^{\alpha_i{I}}\rbrace$
and  ${\tilde{L}}^{\alpha_i{I}}$ is an on-shell operator.
Unfortunately for $n>3$ the expressions for the gauge
symmetry generators are increasingly cumbersome
and the manifest construction of the generalized $C$-ghosts
becomes complicated.
With the gauge symmetry generators (35), (39),
the generalized $B$-ghosts (37)
in the adjoint of the gauge symmetry group
  and the generalized $C$-ghosts (42) - (45)
it is straightforward to construct
the nilpotent BRST charges related to these gauge symmetries.
By definition, the BRST charge is given by
\eqn\grav{\eqalign{
Q=\sum_{n,{\underline{N}}}C_n^{{\underline{N}}}T_{n}^{{\underline{N}}}
+{1\over2}\sum_{m,n,{\underline{M}},{\underline{N}},{\underline{P}}}
(m-n)f^{{\underline{M}}{\underline{N}}}_{{\underline{P}}}
C_m^{{\underline{M}}}C_n^{{\underline{N}}}B_{n+m}^{{\underline{P}}}}}
where
$C_n^{{\underline{N}}}=\oint{{dz}\over{2i\pi}}{z}^{n-1}C^{{\underline{N}}}$
and $B_n^{{\underline{N}}}=n!\oint{{dw}\over{2i\pi}}{1\over{w^{n+1}}}
b_{-1}{\tilde{L}}^{{\underline{N}}}$

The manifest expression for BRST charge is particularly simple
if we restrict ourselves to the operators of the first non-trivial
cohomology $R_2$. In this case there are 2 commuting operators
with identical structure
${\tilde{L}}^{\alpha_{1}\pm}$:
\eqn\grav{\eqalign{{\tilde{L}}^{\alpha+}(w)
={1\over2}\oint{{dz}\over{2i\pi}}(z-w)^2{\lbrack}{1\over2}e^\phi
{F}(X,\psi)P^{(2)}_{2\phi-2\chi-\sigma}(z)+
\cr
4c\xi(F(X,\psi)G-{1\over2}L(X,\psi)P^{(2)}_{\phi-\chi}
-{1\over4}\partial{L}(X,\psi)
P^{(1)}_{\phi-\chi})-24\partial{c}{c}e^{2\chi-\phi}F(X,\psi)
\rbrack}}
\eqn\grav{\eqalign{{\tilde{L}}^{\alpha-}(w)
={1\over2}\oint{{dz}\over{2i\pi}}(z-w)^2{\lbrack}{1\over2}e^\phi
{F}(\varphi,\lambda)P^{(2)}_{2\phi-2\chi-\sigma}(z)+
\cr
4c\xi(F(\varphi,\lambda)G_L-{1\over2}L(\varphi,\lambda)P^{(2)}_{\phi-\chi}
-{1\over4}\partial{L}(\varphi, \lambda)
P^{(1)}_{\phi-\chi})-24\partial{c}{c}e^{2\chi-\phi}F(\varphi,\lambda)
\rbrack}}
where, as previously,
\eqn\grav{\eqalign{
F(X,\psi)\equiv{F_{{5\over2}}}
=\psi_m\partial^2{X}^m-2\partial\psi_m\partial{X^m}\cr
L(X,\psi)\equiv{L_2}=2\partial\psi_m\psi^m-\partial{X_m}\partial{X^m}\cr
G=-{1\over2}\psi_m\partial{X^m}}}
and
\eqn\grav{\eqalign{
F(\varphi,\lambda)=\lambda\partial^2\varphi-2\partial\lambda\partial\varphi\cr
L(\varphi,\lambda)=2\partial\lambda\lambda-(\partial\varphi)^2\cr
G_L=-{1\over2}\lambda\partial\varphi}}
where $\phi$ is the Liouville coordinate (or a coordinate for
 $S^1$ compactified direction in the critical $d=10$ case)
and $\lambda$ is superpartner of $\varphi$.
In uncompactified critical case (which is particularly of interest to
us in this work), the only generator in $R_2$ is ${\tilde{L}}^{\alpha+}$,
while all higher cohomologies are empty.
In this case, ${\tilde{L}}^{\alpha+}(w)$
gives rise to 2 commuting gauge symmetry generators
$L_m{\alpha+}=\partial^m{\tilde{L}}^{\alpha+}(w)(m=1,2)$
which makes the expression (46) for the BRST charge remarkably simple,
so that it can be written as a  worldsheet integral
\eqn\lowen{
Q_1=\oint{{dz}\over{2i\pi}}{\lbrace}ce^\phi{F(X,\psi)P^{(1)}_{\phi-\chi}}
-{1\over8}e^{2\phi-\chi}(L(X,\psi)P^{(2)}_{2\phi-2\chi-\sigma}+2GF(X,\psi))
-\partial{c}c\xi{L}(X,\psi)\rbrace(z)}
The nilpotent charge for the compactified or the noncritical case can be 
obtained by replacing $F(X,\psi)\rightarrow{F}(X,\psi)+F(\varphi,\lambda)$
and $L(X,\psi)\rightarrow{L}(X,\psi)+L(\varphi,\lambda)$
along with $G=G+G_L$,
where $G_L=-{1\over2}\lambda\partial\phi+{q\over2}\partial\lambda$
is the supercurrent in super Liouville theory
($q$ is the Liouville background charge, which of course is
absent in critical  compactified case).
The BRST charge $Q_1$ is the element of superconformal ($\beta-\gamma$) 
ghost cohomology
$H_1$; it reflects the gauge symmetries originating from the
$\alpha$-generators of $R_2$.
Analogously, unifying the alpha-generators of $R_2$ and $R_4$ and using (46)
one can construct the BRST charge $Q_{R_2,R_4}$
 related to the guge symmetries originating
from the first two $b-c$ cohomologies $R_2$ and  $R_4$.
As it turns out that  $Q_{R_2,R_4}$ commutes with $Q_1$,
one can define the nilpotent BRST charge 
\eqn\lowen{Q_2= Q_{R_2,R_4}-Q_1}
reflecting the gauge symmetries originating from $R_4$.
The obtained nilpotent charge $Q_2$ turns out to be the element of
the superconformal $\beta-\gamma$ ghost cohomology $H_2$.  
Next,
 unifying the alpha-generators of $R_2$, $R_4$,  $R_6$ and using (46)
one can construct the BRST charge $Q_{R_2,R_4, R_6}$
 related to the guge symmetries originating
from the first three $b-c$ cohomologies $R_2,R_4$ and $R_6$
As it turns out that  $Q_{R_2,R_4,R_6}$ commutes with $Q_1$ and $Q_2$
one can define the nilpotent BRST charge 
\eqn\lowen{Q_3= Q_{R_2,R_4,R_6}-Q_1-Q_2}
reflecting the gauge symmetries originating from $R_6$.
The obtained nilpotent charge $Q_2$ turns out to be the element of
the superconformal $\beta-\gamma$ ghost cohomology $H_3$. 
In principle, the construction could be continued
to higher values of $n>3$ to construct
the sequence of nilpotent BRST charges $Q_n$  reflecting the
gauge symmetries associated with the $b-c$ cohomologies $R_{2n}$
of higher ghst numbers. However, as it was mentioned above,
the  expressions for the gauge symmetry generators associated
with the $\alpha$-transformations  of $R_{2n}$ become increasingly
complicated for $n>3$ and at this stage it seems
 hard to deduce
any manifest expressions for $Q_n$ at $n>3$.
The chain of the BRST charges (46) generally defines the RNS theories
(as well as kinetic terms of appropriate string field theories)
at various space-time backgrounds.
The form of these backgrounds is defined by 
the geometry of the extra dimensions
induced by the underlying $global$ $\alpha$-symmetry generators
${\tilde{L}}^{\alpha_n{I}}$. Exploration of properties
of these backgrounds (typically of the $AdS_m\times{CP_n}$
or $AdS_m\times{S^n}$ type is an interesting problem which is currently
under investigation ~{\progress}. In the next section
we shall return to the pure spinor formalism,
demonstrating the map between $Q_1$ and the deformed pure spinor
BRST operator in theory with the  nilpotent charge (52) and the
leading singularity order of $(z-w)^{-16}$
in OPE $\lambda(z)\lambda(w)$
of two deformed pure spinor variables.
In the concluding section we shall attempt to extrapolate our results to relate 
the higher order $Q_n$ charges (53), (54), (47) in RNS
approach to deformed pure spinors with higher order OPE singularities.

\centerline{\bf 3. Sequence of New BRST Charges in RNS Theory
and Deformed Pure Spinors}

In this section we discuss the relation between the sequence
of BRST charges described in the previous section and deformed
pure spinor theories outlined in section 1.
The results of this section (as well as the discussion
in the concluding section) apply to $critical$ ten-dimensional
superstring theories (with one of the dimensions
possibly compactified on $S^1$)
We already have pointed out
 that the standard BRST charge $Q_0$ of the RNS theory
can be obtained (up to similarity transformation) from the
deformed pure spinor $BRST$ charge with $\lambda$'s having
a double pole OPE singularity, using the RNS representation
(5) for $\lambda^\alpha$ and (9) for $d_\alpha$.
The goal now is to find  an interpretation for the 
alternative BRST charges $Q_n (n=1,2,3,...)$ in RNS approach
 in terms of the deformed pure
spinors theories with BRST charge having the form (1) but
with the different RNS representations of $\lambda$
(and hence different OPE structures).
Namely, we shall look for the RNS representations
of $\lambda$'s which are commuting space-time spinors and
 dimension zero primary fields, satisfying the relaxed pure spinor
condition - that is, the condition (2) fufilled up to terms with
vanishing normal ordering with  the appropriate OPE
terms of $d_\alpha(z)d_\beta(w)$.
For that we shall require that $\lambda$  still can
be written BRST commutator:
${\lambda^\alpha}=\lbrace{Q_0},\theta^\alpha\rbrace$
with $\theta^\alpha$ still being a dimension 0 primary field and
an anticommuting
space-time spinor  but at higher superconformal ghost numbers
(not related to the ghost  number $1\over2$ expression
(6) by any picture-changing transformation).
We start with the ghost number ${3\over2}$. The natural
way of obtaining a ghost number $3\over2$ anticommuting spinor,
unrelated to the standard ghost number $1\over2$ RNS representation
of the Green-Schwarz variable $\theta$ is to act on (6) with the
truncated 
 $\alpha$-generator of minimal ghost number 1:
 
\eqn\grav{\eqalign{{\tilde{\theta}}^\alpha(z)
={\lbrack}L^{\alpha_1+},\theta^\alpha\rbrack
\equiv\lbrack{\oint{{dw}\over{2i\pi}}}e^\phi(\psi_m\partial^2{X^m}
-2\partial\psi_m\partial{X^m})(w),e^{{1\over2}\phi}\Sigma^\alpha(z)\rbrack
\cr=
e^{{3\over2}\phi}{\tilde{\Sigma}}^\beta\gamma^m_{\alpha\beta}
(2\partial^2{X_m}+\partial{X_m}\partial\phi)(z)}}
(as previously, $\Sigma,{\tilde{\Sigma}}$
are the space-time spinors of opposite GSO parities)
 The ${\tilde{\theta}}^\alpha$ field is a dimension 0
primary field and a
space-time spinor at ghost number ${3\over2}$, {\it not} related 
to the ghost number ${1\over2}$ version (6) of the Green-Schwarz variable by 
picture-changing.
Next, one defines the new ``pure spinor'' variable
${\tilde{\lambda}}^\alpha$ as the BRST commutator
of $Q_0$ with ${\tilde{\theta}}^0$:
\eqn\grav{\eqalign{
{\tilde{\lambda}}^\alpha=\lbrace{Q_0},{\tilde{\theta}}^\alpha
\rbrace}}
The evaluation of the commutator is quite straightforward giving the answer
\eqn\grav{\eqalign{{\tilde{\lambda}}^\alpha=
{\tilde{\lambda}}^\alpha_1+{\tilde{\lambda}}^\alpha_2
+{\tilde{\lambda}}^\alpha_3+{\tilde{\lambda}}^\alpha_4
+{\tilde{\lambda}}^\alpha_5}}
where
\eqn\grav{\eqalign{{\tilde{\lambda}}^\alpha_1
=ce^{{3\over2}\phi}\gamma_m^{\alpha\beta}{\lbrack}
\partial{\tilde{\Sigma}}_\beta(2\partial^2{X^m}+\partial{X^m}\partial\phi)\cr
+{\tilde{\Sigma}}_\beta(2\partial^3{X^m}+4\partial^2{X^m}\partial\phi
+\partial{X^m}({3\over2}(\partial\phi)^2+\partial^2\phi))\rbrack}}
\eqn\grav{\eqalign{{\tilde{\lambda}}^\alpha_2
=-{1\over2}\gamma_m^{\alpha\beta}e^{{5\over2}\phi-\chi}\Sigma^\lambda
\lbrace(2\partial^2{X^m}+\partial{X^m}\partial\phi)\lbrack
{(\gamma_n)_{\beta\lambda}}(2\partial^2{X^n}+\partial{X^n}P^{(1)_{\phi-\chi}})
\cr
+{(\gamma_n\gamma_{pq})_{\beta\lambda}}\partial{X^n}\psi^p\psi^q\rbrack\cr
+\partial{X^m}{\lbrack}{(\gamma_n)_{\beta\lambda}}
({1\over2}P^{(2)}_{\phi-\chi}\partial{X^n}
+P^{(1)}_{\phi-\chi}\partial^2{X^n}
\cr
+{1\over2}\partial^3{X^n}
+{1\over2}\partial{X^n}(\partial\psi_s\psi^s))
+{(\gamma_n\gamma_{pq})_{\beta\lambda}}\psi^p\psi^q
(2\partial^2{X^n}+\partial{X^n}P^{(1)_{\phi-\chi}})\rbrack\rbrace}}

\eqn\grav{\eqalign{{\tilde{\lambda}}^\alpha_3=
e^{{5\over2}\phi-\chi}\Sigma^\alpha\lbrace
{5\over6}P^{(4)}_{\phi-\chi}-{2\over3}\partial\phi{P^{(3)}_{\phi-\chi}}
+{\partial\psi_s\psi^s}(-{5\over8}{P^{(2)}_{\phi-\chi}}
-2\partial\phi{P^{(1)}_{\phi-\chi}})\cr
+{\partial^2\psi_s\psi^s}
{P^{(1)}_{4\phi-5\chi}}-{5\over4}\partial^3\psi_s-
{{115}\over{97}}\partial^2\psi_s\partial\psi^s\rbrace}}
\eqn\grav{\eqalign{{\tilde{\lambda}}^\alpha_4
=e^{{5\over2}\phi-\chi}\Sigma^\lambda{(\gamma_{pq})_{\alpha\lambda}}
\lbrace\psi^p\psi^q({{10}\over3}{P^{(3)}_{\phi-\chi}}-2\partial\phi
{P^{(2)}_{\phi-\chi}})
\cr
+\partial\psi^p\psi^q(10{P^{(2)}_{\phi-\chi}}
-4\partial\phi{P^{(1)}_{\phi-\chi}})+{5\over3}
\partial^2\psi^p\psi^q{P^{(1)}_{\phi-\chi}}
\cr
+5\partial^3\psi^p\psi^q-{{25}\over3}\partial^2\psi^p\partial\psi^q
-{{20}\over3}\partial\psi_s\psi^s\partial\psi^p\psi^q\rbrace}}
\eqn\grav{\eqalign{{\tilde{\lambda}}^\alpha_5
=-{1\over8}be^{{7\over2}\phi-2\chi}\gamma_m^{\alpha\beta}
{\tilde{\Sigma}}^\beta(2\partial^2{X^m}+\partial{X_m}\partial\phi)
P^{(2)}_{2\phi-2\chi-\sigma}}}
The next step is the evaluation of the normal ordered
product $:{\tilde{\lambda}}^\alpha{d_\alpha}:$.
The calculation is quite lengthy but very similar
to the one performed in ~{\psrns} for the $Q_0$ case
with the standard $\lambda^\alpha$  and $d_\alpha$ given by (5), (9).
The result is given by
\eqn\grav{\eqalign{:{\tilde{\lambda}}^\alpha{d}_\alpha:(z)=
{\lbrace}ce^\phi{F(X,\psi)P^{(1)}_{\phi-\chi}}
-{1\over8}e^{2\phi-\chi}(L(X,\psi)P^{(2)}_{2\phi-2\chi-\sigma}+2GF(X,\psi))
\cr
+{7\over2}\partial{c}c\xi{L}(X,\psi)(z)\rbrace}}
This looks similar to the integrand of the 
$Q_1$ BRST charge, except for the last term, having
 the normalization different from (52).
It isn't difficult to see, however, that the worldsheet
integral of $:{\tilde{\lambda}}^\alpha{d}_\alpha:$ is different
from  $Q_1$ just by a similarity transformation.
Namely, it isn't difficult to check that
\eqn\lowen{\lbrack{Q_1},\partial{c}c\partial^2\xi\xi{e}^{-2\phi}\rbrack
=-{9\over8}\partial{c}c\xi{L}(X,\psi)}
Writing
\eqn\lowen{\oint{{dz}\over{2i\pi}}{\tilde{\lambda}}^\alpha{d}_\alpha
=Q_1+{9\over2}\oint{{dz}\over{2i\pi}}\partial{c}c\xi{L}(X,\psi)}
it is easy to see that
\eqn\lowen{\oint{{dz}\over{2i\pi}}{\tilde{\lambda}}^\alpha{d}_\alpha
=e^{-R_1}Q_1{e^{R_1}}}
with
\eqn\lowen{R_1=-4\oint{{dz}\over{2i\pi}}\partial{c}{c}\partial^2\xi
\xi{e^{-2\phi}}}
The nilpotence of $Q_1$ ensures that the deformed pure spinor variable
${\tilde{\lambda}}^\alpha$ satisfies the ``weak'' pure spinor condition
(that is, up to the terms  with the vanishing normal ordering with
$\Pi_m$). It has a singular OPE with itself, with
the leading singularity order equal to $16$:
\eqn\lowen{{\tilde{\lambda}}_\alpha(z){\tilde{\lambda}}_\beta(w)
\sim{{\gamma^m_{\alpha\beta}\partial{b}b{e^{7\phi-4\chi}}\psi_m(w)}\over
{(z-w)^{16}}}+...}
This defines the map of the $BRST$ charge
$Q_1$ (related to the gauge symmetries derived from the
$R_2$ BRST cohomology) to deformed pure spinor theory
with the leading OPE singularity order of $16$ and presents
the main result of this work.
In the concluding section we will discuss possible generalizations
of this result relating the RNS charges $Q_n$ 
derived from higher $b-c$ cohomologies
$R_{2n}$ and  interacting pure spinors with higher order singularities.

\centerline{\bf 5. Discussion and Conclusion}

In this paper we particularly have constructed the map between one of 
the recently
derived BRST charges ~{\sequence}
( related to local ghost matter mixing
gauge symmetries in RNS formalism) and the 
BRST charge  (66) with the ${\tilde{\lambda}}$ variables subject to relaxed
pure spinor constraint and with singular operator products.
The ${\tilde{\lambda}}$ variables have been obtained as 
BRST anticommutator of  the standard 
BRST charge $Q_0$ with the Green Schwarz variables
transformed by the truncated $\alpha$-symmetry generators of ghost number
$1$. The natural question is whether the construction can be generalized
to relate the higher order BRST charges $Q_n$
to deformed pure spinors with higher order OPE singularities.
The natural way to extend the construction demonstrated in this paper
is to
start with the transformations of the Green-Schwarz 
variables $\theta_\alpha$ with truncated $\alpha$-generators
of minimal ghost number $n$:

\eqn\lowen{\theta_\alpha\rightarrow{\theta}_\alpha^{{\underline{N}}}
=\lbrack{L^{{\underline{N}}},\theta_\alpha\rbrack}}
where ${\underline{N}}={\alpha_n{I}};I\equiv(\alpha_{n-1},...,\alpha_1)$
labels the indices of the  truncated $\alpha$-symmetry generators
(obviously excluding the generators that are the Lorenz vectors)
at minimal ghost number $n$. Here and below we shall use
the underlined index notation for the generators in order to avoid any 
confusion with 
space-time fermionic index $\alpha$.
Next, one constructs the commuting spinor variables
(primary field of conformal dimension zero) as
\eqn\lowen{\lambda_\alpha^{{\underline{N}}}=
\lbrace{Q_0},{{\theta}}_\alpha^{{\underline{N}}}\rbrace.}
Then, for each ghost number $n$ one has to find
certain linear combination of the 
commuting spinor variables
$\lambda_\alpha^{(n)}=\sum_{I(n)}\lambda_\alpha^{{\underline{N}}(n)}$
where ${\underline{N}}(n)\equiv(\alpha_n{I}(n))$
with $I(n)=(\alpha_{n-1}...\alpha_n)$
so the summation is over the $n-1$ different values of the index $I(n)$.
The linear combination has to be constructed so that
the resulting $\lambda_\alpha^{(n)}$ field satisfies the weak pure spinor
constraint (i.e. all the singular
and the normal ordered terms of the OPE
$\lambda_\alpha^{(n)}(z)\lambda_\beta^{(n)}(z)$ have vanishing
normal ordering with the relevant terms
of the OPE $d^\alpha(z)d^\beta(w)$. If the conjectured correspondence
between the deformed pure spinor theories and the RNS models defined
by the $Q_n$ BRST charges is correct for $n>1$,
 such a combination should exist and should be
unique for each $n$.
Finally, one constructs the charge
\eqn\lowen{Q_{n}^{PS}=\oint{{dz}\over{2i\pi}}
\lambda_\alpha^{{(n)}}d^\alpha}
by computing the normal ordered product of $\lambda_\alpha^{(n)}$
with $d^\alpha$ of (9).
The constructed charges, written in terms of the $RNS$ variables
should presumably reproduce the $Q_n$ charges in the $RNS$ formalism
related to the $R_{2n}$ cohomologies; however, due to the complexity of 
all the relevant expressions beyond $n=1$ we haven't been able so far to 
prove such a correspondence for higher values of $n$.
The leading term in the OPE of two
$\lambda_{{(n)}}$ 's is given by
\eqn\lowen{\lambda_\alpha^{{(n)}}(z)\lambda_\beta^{{(n)}}(w)
\sim{{{\gamma^m_{\alpha\beta}}\partial{b}be^{(2n+5)\phi-4\chi}\psi_m(w)}
\over{(z-w)^{-2(n^2+6n+1)}}},}
so the correspondence between deformed pure spinors
and $R_{2n}$ related BRST charges in RNS formalism 
(if it holds) implies  that  orders  of $R_{2n}$ cohomologies
giving rise to $Q_n$ BRST charges in RNS formalism, should be in one to one
 correspondence with the leading singularity orders of the OPE 
of deformed pure spinors, given by $2n^2+12n+2$.

\listrefs

\end